\documentclass[letterpaper]{article}

\usepackage[T1]{fontenc}

\usepackage{geometry}
\usepackage{setspace}

\usepackage[style=chem-acs,articletitle=true]{biblatex}
\usepackage{graphicx}
\usepackage{float}
\newfloat{scheme}{htbp}{los}
\floatname{scheme}{Scheme}
\floatname{chart}{Chart}
\newfloat{graph}{htbp}{loh}

\usepackage{chemformula} 
\usepackage[version = 4]{mhchem} 

\usepackage{authblk}
\author[1]{Mohamad Bahsoun}
\author[1]{Jesse Groenen}
\author[1]{Gonzague Agez}
\author[1]{S\'ebastien Jouli\'e}
\author[1]{C\'ecile Marcelot}
\author[1]{Robin Cours}
\affil[1]{Universit\'e de Toulouse, CNRS, CEMES, 29 Rue Jeanne Marvig, Toulouse, 31055 Cedex 4, France}
\author[2]{S\'ebastien Kerdiles}
\author[2]{Mathieu Opprecht}
\affil[2]{LETI, CEA-Universit\'e Grenoble-Alpes, 17 Avenue des Martyrs Grenoble, 38054, France}

\author[1]{Caroline Bonafos}
\author[1]{Jean-Marie Poumirol}

\title{Emergence of Localized Surface Plasmons in Unpatterned Hyperdoped Polycrystalline Silicon}

\date{*Email: jean-marie.poumirol@cemes.fr, caroline.bonafos@cemes.fr}

\begin{document}

\maketitle

\begin{abstract}
The ability to engineer localized surface plasmon resonances at large scale usually relies on precise nanoscale patterning. Here, we demonstrate that mid-infrared plasmonic responses can instead emerge in unpatterned polysilicon films composed of nanometric (5–50 nm) grains, challenging established design paradigms and eliminating the need for external nanostructuring. Using tailored out-of-equilibrium annealing conditions, we show that hyperdoped polysilicon layers exhibit enhanced light–matter interactions that can be tuned across the mid-infrared range. By combining advanced electron microscopy, infrared spectroscopy and finite-difference time-domain electrodynamic simulations, we demonstrate that these remarkable optical properties originate from naturally formed metal–dielectric interfaces at grain boundaries, which support localized surface plasmon resonances. Importantly, this result is universal and can be extended to any doped semiconductor system, regardless of the synthesis technique, provided that the grain size remains in the nanometric range. This work opens up a new field in plasmonics centered on polycrystalline semiconductors, paving the way for cost-effective systems that are fully compatible with microelectronic and photovoltaic technologies, and capable of significantly reshaping light–matter interactions in the infrared range.

\end{abstract}

\section*{Keywords}
Infrared plasmonics, enhanced light matter interaction, polysilicon



\section{Introduction}\label{sec1}
Advances in nanofabrication enabling subwavelength metasurfaces have been central to the success of plasmonics \cite{yang_plasmonic_2025, wang_plasmonic_2016, hu_review_2021, cortes_optical_2022, meinzer_plasmonic_2014}. However, despite rapid progress, the widespread commercialization of plasmonic metasurfaces remains hindered by challenges in scalable manufacturing and deployment in real-world devices \cite{patoux_challenges_2021}. Even if significant efforts have been made to overcome the inherent limitations of electron-beam lithography - the dominant fabrication method for metasurfaces - there is still a strong need for cost-effective and environmentally friendly mass manufacturing of optical metasurfaces \cite{seong_cost-effective_2024}. 
The recent emergence of localized surface plasmon resonances (LSPR) in doped semiconductor nanocrystals represents a paradigm shift, replacing traditional noble-metal nanostructures with more sustainable solutions, extending plasmonics into the infrared while addressing the environmental and supply risks associated with critical metals \cite{graedel_criticality_2015}.  Among semiconductors,
silicon (Si) is of particular interest due to its low production cost, abundance, and non-toxicity. These semiconducting nanostructures fundamentally differ from their metallic counterparts due to their lower but widely tunable free-carrier concentrations. \cite{luther_localized_2011, naik_alternative_2013, law_towards_2013, faucheaux_plasmon_2014, kriegel_plasmonic_2017, taliercio_semiconductor_2019} The plasma frequency can then be freely tuned via impurity doping allowing LSPR engineering, giving an additional degree of freedom compared to metals, which have a high and fixed electron density. This new way of controlling the resonance frequency combined with the resulting expansion of plasmonics into the infrared (IR) spectral range, not easily reached with metals, has fueled the search for such novel plasmonic materials with improved crystalline quality, integrability, tunability, stability, and potentially lower losses \cite{faucheaux_plasmons_2013, zhou_comparative_2015, kramer_plasmonic_2015, pellegrini_benchmarking_2018, poumirol_hyper-doped_2021, valdenaire_heavily_2023, zhang_infrared_2022, zhang_hyperdoped_2023}. However, achieving light–matter interaction to application-relevant levels requires a large-scale implementation of plasmonic metastructures based on highly organized and densely packed nanometric antennas, while maintaining high electron mobility \cite{poumirol_hyper-doped_2021, baldassarre_midinfrared_2015, taliercio_semiconductor_2019}.\\

\begin{figure*}
    \centering
    \includegraphics[width=1\linewidth]{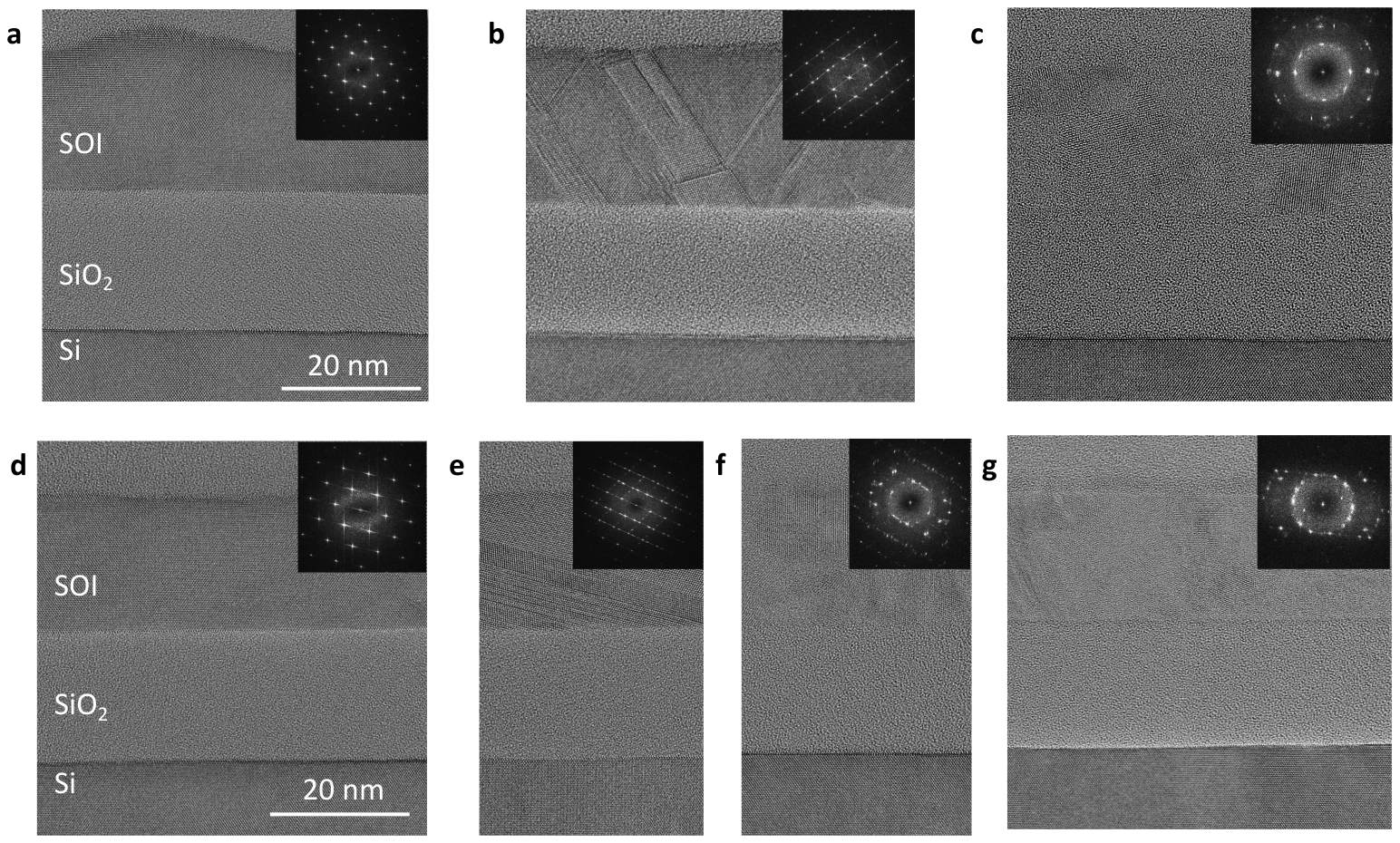}
    \caption{\textbf{Microstructural evolution at the melting threshold}. High Resolution Transmission Electron Microscopy (HRTEM) images in cross-section of samples D1 and D2 after LTA: before Full Melt (FM)(a) and (d), at FM (b) and (e),(f), after FM (c) and (g). In inset, associated Fast Fourier Transforms of the images. The laser energy densities at FM for D1 and D2 are given in Table S1(SI). For the two doses, the hyperdoped Si top layer is monocrystalline below FM and polycrystalline beyond FM. At the FM onset, highly twinned silicon layers are obtained, together with polycrystalline regions for D2.}  

    \label{fig:1}
\end{figure*}

Here we introduce an approach to realizing large-area, ultrathin silicon-based plasmonic layers without engineered nanostructures. We show that heavily doped continuous polycrystalline silicon thin films exhibit plasmonic behavior that differs fundamentally from that of conventional metallic films, with the ability to intrinsically sustain localized surface plasmon resonances. By combining transmission electron microscopy (TEM), Fourier-transform infrared spectroscopy and electrodynamic simulations, we uncover that these unexpected properties stem from the intrinsic microstructure of hyperdoped polysilicon, where metallic-like silicon nanograins are surrounded by ultrathin insulating grain boundaries. These spontaneously formed, densely packed metal–dielectric interfaces provide a framework for enhanced light–matter interaction in the infrared, offering a viable strategy for creating low-cost, sustainable plasmonic materials that bypass complex nanofabrication steps while remaining fully compatible with microelectronic and photovoltaic platforms.
\section{Results}

\begin{figure*}[t]
    \centering
    \includegraphics[width=1\linewidth]{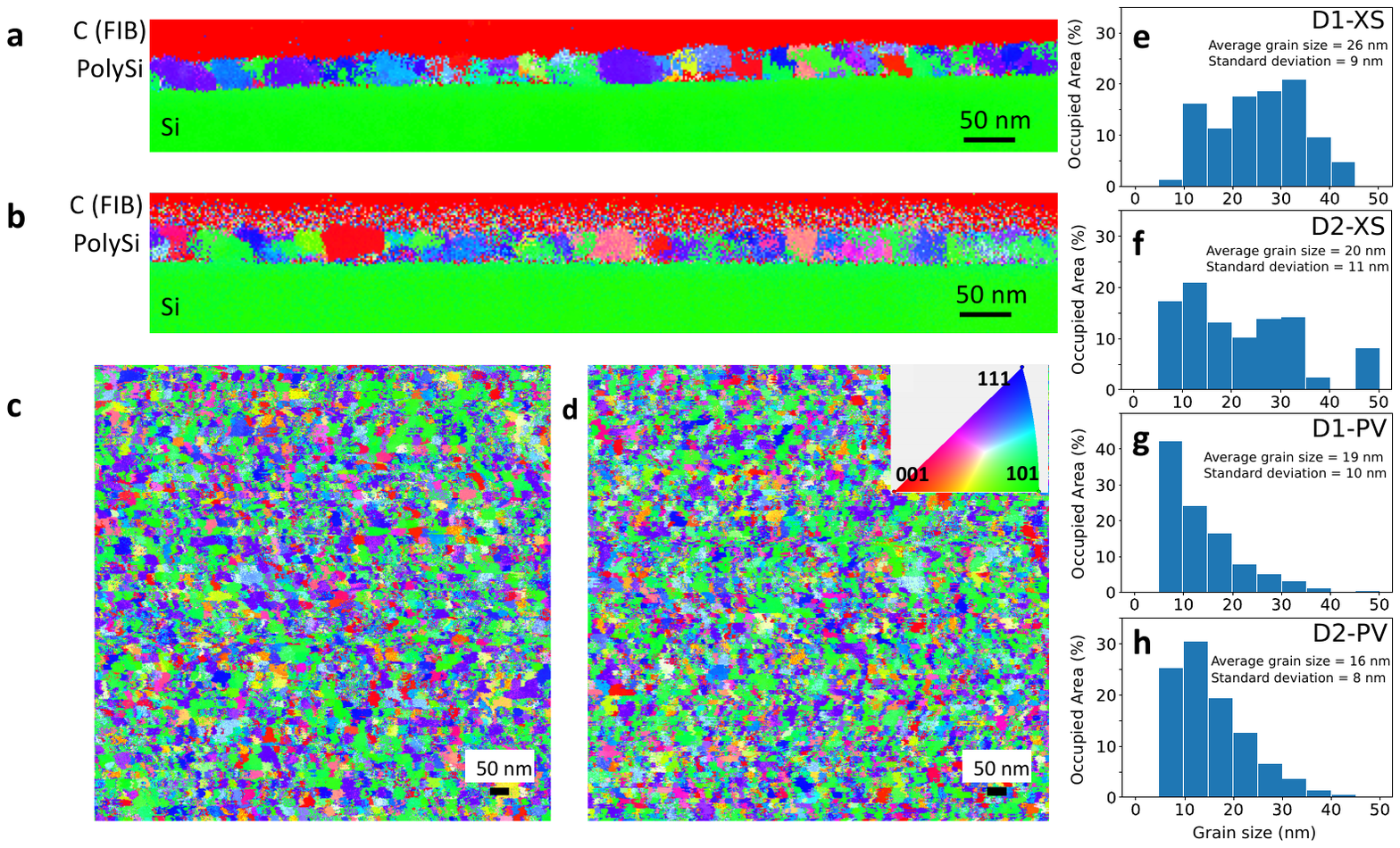}
    \caption{\textbf{Orientation maps and grain-size distributions of polysilicon layers.} ACOM images (orientation maps) of D1 ((a),(c)) and D2 ((b), (d)) annealed beyond the FM threshold, in cross-section (XS) and plan-view (PV); (e), (f), (g) and (h) associated size-distributions. The hyperdoped polycrystalline layer consists of randomly oriented nanograins with sizes ranging between 5 and 50 nm for both D1 and D2.}  
    \label{fig:2}
\end{figure*}
\subsection{Sample Fabrication and structural characterization}
Hyperdoped polysilicon films were obtained by laser thermal annealing (LTA) of a 23 nm phosphorus doped silicon-on-insulator (SOI) substrate (see sample elaboration in SI). In such extremely out-of-equilibrium conditions, concentrations of electrically active dopants beyond the equilibrium solid solubility limit can be achieved in localized regions, while the integrity of the surrounding areas is preserved \cite{huet_doping_2017}. Phosphorus dopants were introduced by low-energy ion implantation at two different doses (labeled D1 and D2, see table S.1, SI). Such high implantation doses partially amorphize ($\approx$ 15 nm) the silicon top layer, leaving a thin crystalline seed at the interface with the SiO$_2$ Buried Oxide (BOx) \cite{chery_study_2022}. LTA restores crystallinity in the as-implanted amorphous layers and enables efficient electrical activation of dopant species. By adjusting the laser energy density, the position of the melt front and thus the crystallinity of the annealed layer can be precisely controlled \cite{chery_study_2022}. When the melt front remains within the crystalline seed—at laser energies just below the full melt (FM) threshold— the regrown layers are fully monocrystalline (Fig. \ref{fig:1} $\textbf{a}$ and $\textbf{d}$). The presence of some hillocks at the surface has been attributed to the late solidification of residual liquid-Si droplets during the recrystallization process \cite{chery_study_2022}. For energies beyond the FM threshold 
, where the melt front reaches the buried oxide, polycrystalline films are formed (Fig. \ref{fig:1} $\textbf{c}$, $\textbf{g}$), due to the lack of an initial crystalline seed. Here, the Fast Fourier Transform shows diffraction rings rather than the discrete reflections of the Si (110) zone axis. At the FM onset, with the melt front pinned at the interface between the crystalline seed and the buried oxide, highly twinned silicon layers are obtained for both D1 (Fig. \ref{fig:1} $\textbf{b}$) and D2 (Fig. \ref{fig:1} $\textbf{e}$), together with some polycrystalline regions for D2 (Fig. \ref{fig:1} $\textbf{f}$). To better visualize the formed polysilicon grains, (Fig. \ref{fig:1} $\textbf{c}$ and $\textbf{g}$) Automated Crystal Orientation Mapping (ACOM) \cite{rauch_automated_2010} in TEM has been performed on the samples annealed beyond full-melt conditions. Under both doping conditions, the polycrystalline layer is composed of randomly oriented nanograins exhibiting a variety of shapes (see Fig. \ref{fig:2} $\textbf{a}$ and $\textbf{b}$ in cross-section and $\textbf{c}$ and $\textbf{d}$ in plan-view). The size distributions are determined from both cross-sectional and plan-view images, where the grain size is defined as the diameter of the surface-equivalent circular grain. In plan-view, the grain size ranges from 5 to 50 nm and follows a log-normal-type distribution (Fig. \ref{fig:2} $\textbf{g}$ and $\textbf{h}$), as commonly reported for polycrystalline layers formed by annealing of amorphous films. This behavior is attributed to the cessation of nucleation resulting from the finite amorphous reservoir \cite{bergmann_formation_1998}. In cross-sectional images, the grain size measured perpendicular to the BOx interface corresponds, for the majority of grains, to the layer thickness (Fig. \ref{fig:2} $\textbf{a}$ and $\textbf{b}$). This geometric constraint likely influences the size distribution, which appears more symmetric (Fig. \ref{fig:2} $\textbf{e}$ and $\textbf{f}$). The microstructure of the polycrystalline layers is very similar for D1 and D2 and the average grain size extracted from plan-view images with robust statistics (Fig. \ref{fig:2} $\textbf{g}$ and $\textbf{h}$) is comparable for the two phosphorus doses (16– 19nm), within experimental uncertainty ($\pm$ 4nm). Therefore, within this range of phosphorus doses, the doping level does not affect the recrystallisation process and the final morphology of the resulting polysilicon layer.

\subsection{Infrared Optical Properties}  

\begin{figure}[h!]
    \centering
    \includegraphics[width=0.5\linewidth]{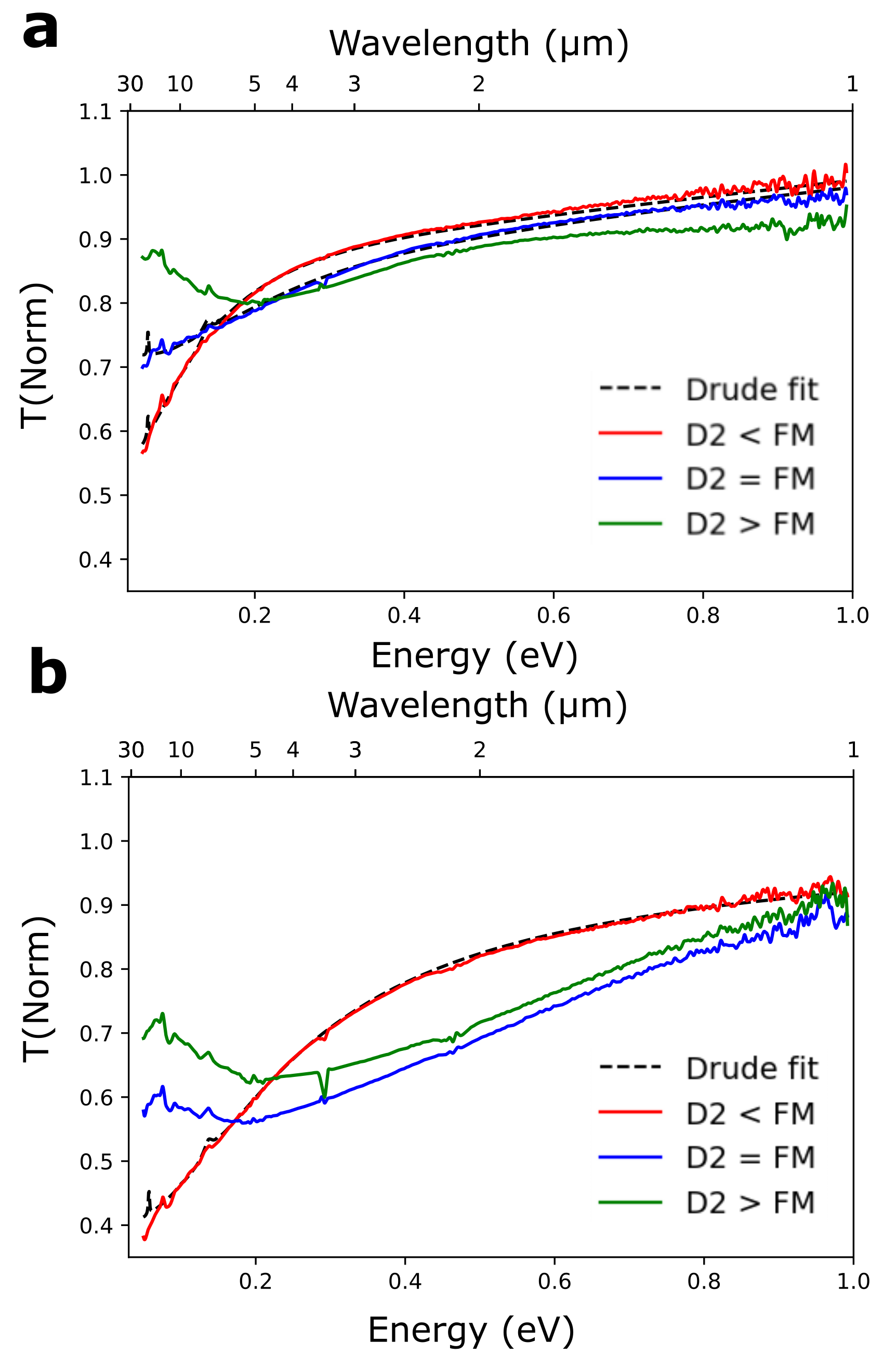}
    \caption{\textbf{Experimental optical response of hyper-doped Si layers.} Normalized transmittance measured for D1 ($\textbf{a}$)  and D2 ($\textbf{b}$). Continuous lines, measured on unpatterned hyper-doped Si layers below full melt, at full melt and beyond full melt. Dashed lines, simulated using model described in the text. For samples annealed below FM, the normalized transmittance remains close to unity at high frequencies and decreases markedly at low frequencies, consistent with free-electron absorption in the n-doped layer (Drude-like response). In contrast, polycrystalline doped Si layers exhibit a pronounced transmission minimum associated with a well-defined resonance that blueshifts with increasing dopant concentration, characteristic of localized surface plasmon (LSP) modes.}  
    \label{fig:3}
\end{figure}

Infrared spectra were recorded at room temperature using a Fourier Transform Infrared (FTIR) spectrometer coupled to a Cassegrain microscope. Figure \ref{fig:3} shows the transmission (T) spectra measured at normal incidence on D1 and D2 samples annealed under the different LTA conditions: below, at, and beyond FM. The spectra were normalized using a reference sample in which the 23 nm-thick doped Si top layer had been completely removed (T=I$_{T(Sample)}$/I$_{T(ref)}$), allowing the contribution of the doped layer to be identified. For samples annealed below FM (red curves in Fig. \ref{fig:3} $\textbf{a}$ and $\textbf{b}$), which consist of a continuous monocrystalline doped Si layer (Fig. \ref{fig:1} $\textbf{a}$ and $\textbf{d}$), the normalized transmission remains close to unity at high frequencies but exhibits a marked decrease at low frequencies. This reduction originates from free-electron absorption in the n-doped layer, consistent with the Drude-like response of a continuous metallic film \cite{poumirol_hyper-doped_2021}. With increasing phosphorus concentration, and thus higher free carrier density, the low-frequency minimum deepens. To extract quantitative information from our optical measurements, we model the optical properties of the multilayer structure: doped Si ($h$ = 23 nm)/Si$O_2$ ($h$ = 20 nm)/intrinsic Si ($h$ = 775 $\mu$m) using the complex refractive index of each layer (see Optical properties of continuous layer in SI). The free carrier density ($N$) and electron scattering time ($\tau$) can then be obtained by fitting the transmission spectra. The corresponding fits, shown as black dashed lines in Fig. \ref{fig:3}, yield N$_{D1}^{<FM}$=3.83$\times$10$^{20}$cm$^{-3}$, $\tau_{D1}^{<FM}$= 5.73$\times$10$^{-15}$s 
for D1 and N$_{D2}^{<FM}$=1.01$\times$10$^{21}$cm$^{-3}$, $\tau_{D2}^{<FM}$=3.4$\times$10$^{-15}$s 
for D2. These values exceed both the solid solubility limit and, more critically, the thermally activated phosphorus concentration in bulk Si at equilibrium \cite{pichler_intrinsic_2004}. As expected D2 that received a higher implantation dose presents higher free carrier concentration. In the D1 sample annealed at the FM (Fig. \ref{fig:3} $\textbf{a}$, blue curve), the transmittance changes slightly, with an apparent broadening and amplitude decrease of the Drude related extinction. For this sample, the fitting procedure gives N$_{D1}^{FM}$=3.81$\times$10$^{20}$cm$^{-3}$, $\tau_{D1}^{FM}$= 2.67$\times$10$^{-15}$s ,
(black dashed line).These values indicate that the extended defects (twin boundaries, Fig. \ref{fig:1} $\textbf{b}$) only reduce carrier mobility, while the activation rate and the resulting free carrier concentration remain unaffected.\\ 

Interestingly, in samples D1 annealed beyond the FM, the shape of the transmission is strongly altered and the transmission minimum is no longer localized at zero frequency but shifted to finite energy ($\omega_{R} \approx$ 5.9 $\mu$m, E$_R \approx$ 208 meV). This is a clear sign that the optical signature is no longer dominated by free carriers. Moreover, as can be seen in Fig. \ref{fig:3} $\textbf{b}$ a similar trend is observed for sample D2 annealed beyond the full melt with a departure from the Drude like behavior. As for D1 beyond FM, 
a minimum in the transmission appears, but at higher energy $\omega_{R} \approx$ 5.5 $\mu$m, E$_R \approx$ 222 meV.

These two samples share the feature of being composed of polycrystalline doped Si layers (see Fig. \ref{fig:1} $\textbf{c}$, $\textbf{g}$ and Fig. \ref{fig:2}). These spectral signatures, reflecting enhanced light–matter interaction at a well-defined resonance that blueshifts with increasing dopant density, are characteristic of LSP resonances, as observed in electron-beam–nanopatterned metasurfaces based on the monocrystalline layers \cite{poumirol_hyper-doped_2021}. The D2 sample annealed at full melt is composed of a mixture of polycrystalline and monocrystalline areas of relatively large size (few $\mu$m$^2$ ) side by side (see Figure \ref{fig:1} $\textbf{e}$ and $\textbf{f}$). The optical response measured in the far field (blue curve in figure \ref{fig:3} $\textbf{b}$) is consistent with the superposition of the two behaviors described above.

The behavior above FM may appear counterintuitive for continuous metallic-like layers as hyperdoped silicon, in which propagating—rather than localized—plasmon polaritons are typically observed \cite{faruque_heavily_2021, shahzad_infrared_2011}. Yet, the polycrystalline layer is composed of randomly oriented nanograins separated by grain boundaries (GBs). In silicon, unlike in plasmonic metals such as gold or silver, GBs are non-metallic and govern electrical transport through dopant segregation and carrier trapping \cite{kim_theory_1984,Kamins1998, mandurah_model_1981, tsurekawa_interfacial_2007}. A substantial fraction of dopants segregates to the GBs, where they become electrically inactive owing to the high density of dangling bonds and structural defects. Consequently, GBs behave as narrow insulating barriers, and charge transport can be described as occurring across heterojunctions between conductive nanocrystals and intrinsically wide-band-gap GB regions \cite{kim_theory_1984,Kamins1998, mandurah_model_1981}. 
In transport models, the effective electrical width of the GBs is typically taken to be 1–2.5 nm, slightly larger than the few-atomic-layer-thick, atomically GB core \cite{ding_tem_2020, sakaguchi_atomic_2007}. The influence of GBs on electrical transport is experimentally evidenced in our samples by a marked increase in resistivity above the FM, where the layers are polycrystalline, compared with the monocrystalline region below the FM, for both D1 and D2 (Fig. S2, Supplementary Information). From an optical viewpoint, assuming that isolating GB can be treated as a dielectric layers, each boundary now forms a metal–dielectric interface capable of sustaining LSP resonances. More broadly, the polysilicon layer can then be described as a dense assembly of metallic-like silicon nanocrystals (5–50 nm) separated by thin dielectric regions.

\begin{figure*}[h!]
    \centering
    \includegraphics[width=\textwidth]{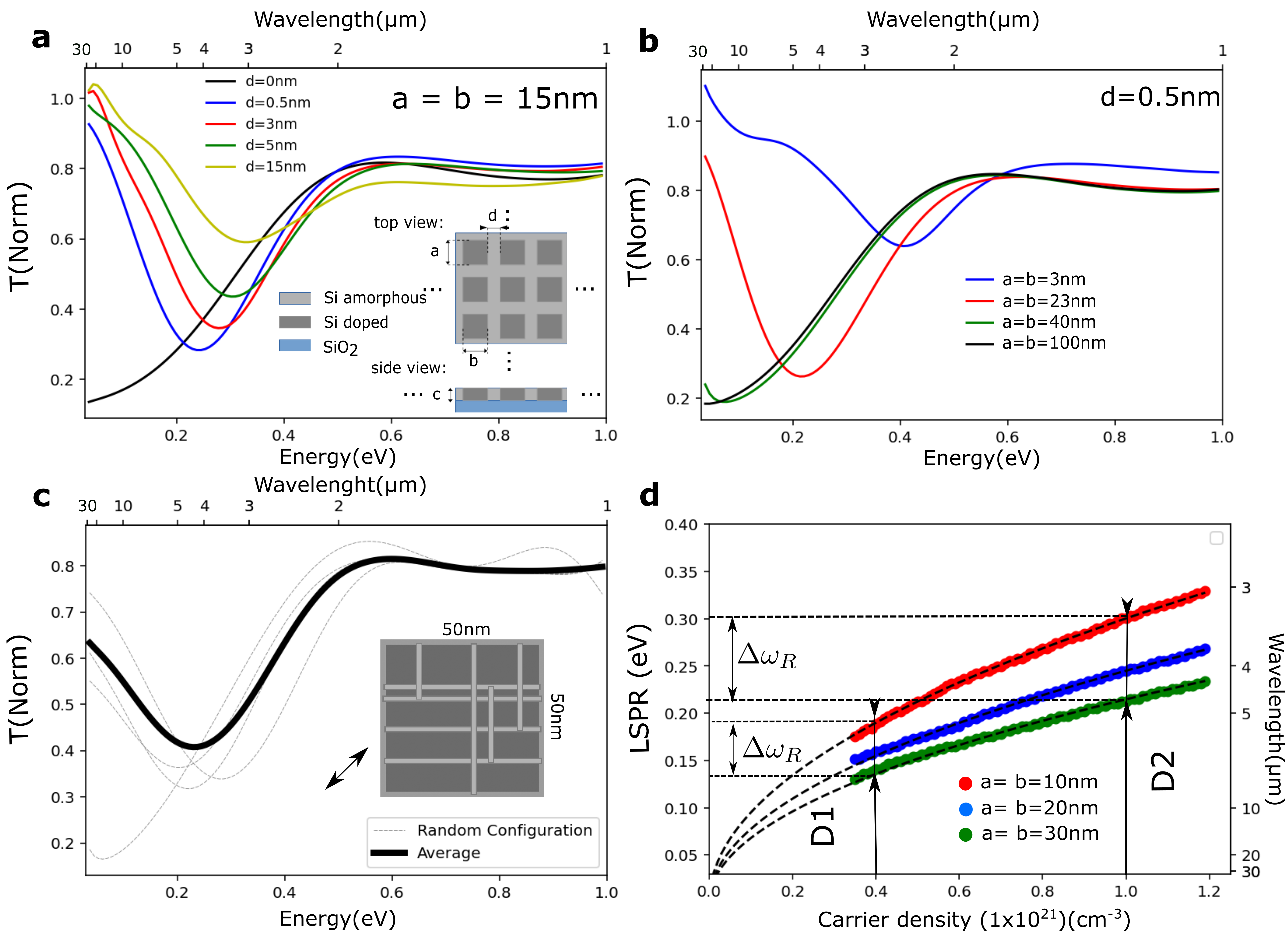}
    \caption{\textbf{FDTD-calculated transmission for doped Si antenna arrays.} $\textbf{a}$. FDTD calculated normalized transmission for doped Si quadratic antenna in infinite square array, with dimension of $a$ =15 nm, $b$ = 15 nm, $c$ = 23 nm with carrier density $N$=1$\times$10$^{21}$cm$^{-3}$ and scattering time $\tau=3.4\times10^{-15}$s. Antennas are separated by a variable thickness $d$ (from 0 nm to 15 nm) of amorphous Si and deposited on SiO$_2$. Insert: schematic configuration used for numerical simulations. Incoming light polarization is fixed as illustrated by the arrow $\textbf{b}$. Calculated normalized transmission as in (a) but for a constant values of $d$ = 0.5 nm and various values of antennas size ($a$ = $b$= 3, 23, 40, 100 nm). $\textbf{c}$. Calculated Transmission for randomized configuration. The antennas lateral dimensions are now randomly selected with $(a, b) \in [2,50] $ nm, and fixed $d$ = 1 nm and the total number and shape of doped antennas are adjusted to ensure that the space is fully covered. Grey lines: Transmission calculated for different random configurations (as illustrated in insert). Black lines: Average transmission obtained for 10 different configurations. $\textbf{d}$. Calculated LSPR as a function of the carrier density for 3 different arrays $a$ = $b$= 10, 20 and 30 nm and fixed $d$ = 5 nm. Dashed lines, fitted square root of $N$ dependence.}  
    \label{fig:4}
\end{figure*} 

\subsection{Numerical simulation} 
To confirm the relevance of this description we performed finite-difference time-domain (FDTD) simulations to calculate the transmission for such metal/dielectric hybrid system (see Numerical simulations in SI). From the perspective of the physical phenomena of interest, hyperdoped polycrystalline Si constitutes a structurally complex system. It combines pronounced disorder with multiple characteristic length scales, ranging from nanometer-wide grain boundaries to nanocrystals extending over several tens of nanometers. Such structural richness lies beyond the reach of fully realistic numerical modeling. We therefore adopt here a simplified geometry designed to capture the dominant physical mechanisms rather than to reproduce the experimental transmission spectra quantitatively. In order to study the effects of grain size and grain boundary, we will first consider ordered systems that are free from fluctuations. We then analyze the effects of disorder and fluctuations. 

The model considers a densely packed array of doped silicon nanometric antennas separated by nanometer-scale dielectric Si regions that define the GBs. As illustrated in figure \ref{fig:4} $\textbf{a}$ using periodic boundaries we simulated hyperdoped silicon quadratic antenna organized in periodic square arrays, the interspace being filled with a dielectric material. Both in-plane antenna dimensions were set equal ($a=b$) and a fixed thickness $c=23$ nm (the nominal thickness of the doped Si layer) is chosen.
The antennas are separated by a distance (gap) $d$. Simulations require defining for each material its dielectric function over the full range of frequency probed. For doped Si antennas, the dielectric function was calculated using the Drude-Lorentz model (see \ref{equa:1} in SI) with a carrier concentration set to $N$=1$\times$10$^{21}$cm$^{-3}$ and a scattering time set to $\tau=3.4\times10^{-15}$s, both derived from the experimental spectrum of D2 
below full melt (Fig. \ref{fig:3} $\textbf{b}$). The dielectric function of the gaps was set to $\epsilon$ = 11.4 which corresponds to amorphous silicon in the frequency range of interest. \cite{oskooi2010meep}

Figure \ref{fig:4} $\textbf{a}$ shows the calculated normalized transmission spectra for periodic square arrays of doped antennas with dimensions $a = b = 15$ nm. We choose these lateral dimensions because it is close to the average grain size measured for D2 (see size histogram in Fig. \ref{fig:2} $\textbf{h}$). The effect of antenna spacing was investigated by varying the gap $d$ between 0 and 15 nm. Each array of antennas exhibits a pronounced transmission minimum. Decreasing $d$ from 15 to 1 nm shifts the resonance to lower frequencies and deepens the transmission minima. When $d$ = 0, one recovers the transmission of a continuous 2D metallic-like layer, with a transmission minimum centered at zero frequency (see black curve in Figure \ref{fig:4} $\textbf{a}$). Those simulations confirm that an array of doped Si antennas even in very close proximity (down to $d$ = 0.5 nm) can sustain LSP resonance that profoundly alters the transmission spectra compared to the continuous layer. The simulation also reveals that the size of the gap $d$ separating antennas is a parameter of prime importance. The strong redshift of the plasmon frequency captured by our numerical simulation is due to the near-field coupling between antennas resulting from the dipole-dipole interaction that occurs when antennas are brought in close vicinity \cite{gunnarsson_confined_2005, rechberger_optical_2003, maier_observation_2002, su_interparticle_2003}.


Figure \ref{fig:4} $\textbf{b}$ displays the normalized transmission spectra calculated for various sizes of square antennas at a constant gap $d$ = 0.5 nm. One can see that the transmission spectra depend strongly on antenna size: increasing antenna size results in a red-shift of the resonance from $\approx$ 2.85 $\mu$m  for $3\times3$ $~\text{nm}^2$ antennas down to $\approx$ 20 $\mu$m  for $40\times40$ $~\text{nm}^2$ antennas. This may seem counterintuitive because the observed resonance wavelengths are much larger than the antenna dimensions ($\lambda/a \in [10^{-2},10^{-3}$]). Consequently, localized surface plasmons are usually described using the quasi-static approximation. In this regime, the plasmon frequency of a single antenna is independent of its size \cite{maier_plasmonics_2007}. However, owing to the strong coupling occurring between adjacent antennas, this approximation is no longer valid. As predicted by our simulations, increasing the antenna size leads to a red shift of the LSP resonance. Our simulations also reveal that increasing further the antenna size (up to $100\times100$ $~\text{nm}^2$) the calculated transmission spectra can no longer be differentiated from the transmission spectra of a continuous doped Si layer. 

Both Figures \ref{fig:4} $\textbf{a}$ and $\textbf{b}$ show that the filling factor, defined as the relative surface occupied by doped silicon $ff= S_{Si}/S_T$, plays an important role on the optical properties of the layer. Indeed, As $ff$ goes from 0.25 (when $d$ = $a$ = $b$ = 15 nm, yellow curve in Figure \ref{fig:4} $\textbf{a}$) to 0.975 (when $a$ = $b$ = 40 nm, $d$ = 0.5 nm, dark green curve in Figure \ref{fig:4} $\textbf{b}$)) the extinction amplitude at resonance frequency is multiplied by a factor 2 (from 40$\%$ up to 80$\%$).

Figures \ref{fig:4} $\textbf{a}$ and $\textbf{b}$ focus on ordered monodisperse antennas. Introducing positional disorder and size fluctuations is known to cause inhomogeneous broadening in optical responses. For very large ensembles, the response is akin to the incoherent sum of single antenna contributions. To illustrate this effect, we performed simulations using randomized configurations. To do so, as illustrated in the inset of Figure \ref{fig:4} $\textbf{c}$, the antennas lateral dimensions are now randomly selected with (a, b) $\in$ [2,50] nm, ranging the grain dimensions observed in our samples. The gap between doped antennas is kept fixed to $d$ = 1 nm and the total number and shape of doped antennas are adjusted to ensure that the space is fully covered, all other parameters (doping, scattering time) being kept identical to previous simulations. The resulting transmission spectra calculated for selected random configurations appear as light grey lines on Figure \ref{fig:4} $\textbf{c}$. All measured configurations sustain LSPR and exhibit a broader resonance, i.e. an increased full width at half maximum (FWHM), compared with the monodisperse simulations, even without modifying the plasmonic scattering time (Figure \ref{fig:4} $\textbf{a}$ and $\textbf{b}$). Furthermore, every configuration yields different plasmon resonance frequencies. The black thick line is the average transmission spectra obtained from 10 randomly selected configurations mimicking the effect of incoherent summing. Because of the dispersion observed for the resonance frequencies, this averaging effect results in an increased FWHM (extrinsic broadening) and a less pronounced extinction.

Finally, we turn our attention to the effect of the carrier density (doping) on such systems. In Figure \ref{fig:4} $\textbf{d}$ we calculate the LSPR frequency for three different size configurations versus carrier density $N$ (fixed $\tau$ and $d$). For any given configuration, the LSPR follows the well known square root dependence with carrier density \cite{maier_plasmonics_2007}. As expected from the previous calculation, at a fixed carrier density and because of the dipole-dipole interaction, the LSPR appears at higher energy for smaller antennas. This simulation also reveals that as the carrier density decreases, the size dependence shift becomes less pronounced (see $\Delta \omega_R (N=1 \times 10^{21}) > \Delta \omega_R (N=0.4 \times 10^{21})$ in Figure \ref{fig:4} $\textbf{d}$). 
As size-dependent LSPR broadening originates from the extrinsic contribution previously described, the measured plasmon FWHM should decrease with decreasing carrier density.

\section{Discussion}
Although based on simplified geometrical descriptions, these simulations provide valuable insight into LSPR within the frequency range relevant to doped polycrystalline silicon. They reveal that LSPR in hyperdoped polysilicon originates from its intrinsic nanostructure, where nanometric metallic domains are naturally surrounded by $\approx$0.5-2 nm dielectric grain boundaries. The plasmonic response is governed by three key structural parameters: the thickness of the dielectric barriers, the degree of grain inhomogeneity (including variations in size and shape) and the carrier density. The sharp dielectric grain boundaries mediate strong dipole–dipole interactions between metallic Si nanodomains, giving rise to a pronounced redshift of the plasmon resonance. This shift has been studied theoretically and experimentally for metallic systems \cite{gunnarsson_confined_2005, rechberger_optical_2003, maier_observation_2002, su_interparticle_2003}. For two neighboring spherical metal nanoparticles the plasmon resonance redshift has been found to depend on one critical parameter, the gap d to diameter D ratio $\Delta\lambda/\lambda \propto $ exp$(-d/D)$ \cite{gunnarsson_confined_2005}. Because the plasmonic doped Si antennas have in average typical dimensions of a few tens of nanometers and are separated by $\approx$1 nm GBs, the dipole–dipole interactions govern the optical properties, explaining why the experimentally measured plasmon appears at such a low frequency even with high carrier concentration.

Remarkably, this strong dipole–dipole interaction also induces a redshift of the LSPR as the antenna size increases, despite the size remaining only $~$10$^{-3}$ of the IR wavelength, well within the quasi-static regime. The intrinsic structural disorder of the polysilicon, composed of grains with varying sizes and shapes (Figure \ref{fig:2} $\textbf{c}$ and $\textbf{d}$) induces an inhomogeneous broadening of the plasmonic response, 
leading to less pronounced extinction minima. As a result, in contrast to ordered nanostructures where Mie theory can be applied, the carrier density cannot be reliably extracted from the LSP resonance frequency. Nevertheless, in very good agreement with the experimental observations the simulations show that decreasing the carrier concentration leads to lower resonance frequency and to a decrease of extrinsic broadening. Therefore, in this system, carrier density remains a relevant tuning parameter to control the plasmonic based optical properties.  

At last, the simulations yield valuable insight into the physical conditions under which a polycrystalline system can support LSP resonances. Given $\approx$ 0.5 to 3 nm grain-boundary separations and dominant dipole–dipole coupling, an overly large average nanocrystal size would shift the plasmon resonance so strongly toward lower energies that the optical response of the polycrystalline layer would be indistinguishable from the Drude-like behavior observed for monocrystalline Si. 

Finally, these plasmonic layers combine two distinctive strengths. The first is their dramatically high filling factor (average $ff$ = 0.88), arising from nanometric thin dielectric grain boundaries, maximizes light–matter interaction and enables strong optical confinement within an ultrathin medium (22 nm). Transmission measurements reveal that the polysilicon layer extincts at the plasmon resonance frequency a very substantial fraction of the incident light, with extinction reaching ~20$\%$ for D1($>$FM) and up to ~37$\%$ for D2($>$FM) (green curves of Fig. \ref{fig:3} $\textbf{a}$ and $\textbf{b}$), which is particularly remarkable given the extremely small thickness of the active layer. These values largely exceed those reported for metasurfaces based on nanopatterned monocrystalline layers (red and blue curves of Fig. 3 $\textbf{b}$) from \cite{poumirol_hyper-doped_2021}, which exhibited only $\approx$8$\%$ absorption despite a relatively high 0.4 filling factor) and for plasmonic systems based on hyperdoped silicon nanoantennas embedded in silica \cite{valdenaire_heavily_2023, zhang_infrared_2022, zhang_hyperdoped_2023} at similar optical frequencies. 
 
The second key strength resides in the spontaneous formation of these hyperdense nanometric arrays over large areas, without requiring any post-growth nanopatterning. This intrinsic scalability, together with their robust plasmonic response, establishes hyperdoped polysilicon as a compelling platform for integrated and large-area infrared nanoplasmonics. Notably, this approach is not limited to polysilicon but can be generalized to any doped semiconductor system with nanometric grain sizes, regardless of the fabrication method. Furthermore, unlike top-down metasurfaces fabricated via electron-beam lithography or other high-resolution techniques, these nanocrystalline layers combine exceptional robustness with large-scale uniformity and can be deposited on arbitrary surfaces, including flexible or non-planar substrates.

\section{Conclusion}
In conclusion, we uncover a paradigm shift in plasmonics, demonstrating that localized surface plasmon resonances in the mid-infrared can emerge intrinsically from unpatterned polysilicon layers composed of nanometric (5–50 nm) grains, obviating the need for any external nanostructuring. This behavior, which cannot be achieved in polycrystalline noble metals, arises from the strong dielectric contrast between metallic-like silicon grains and dielectric grain boundaries. Owing to the nanometric thickness of the grain boundaries, the system reaches an exceptionally high filling factor, enabling intense light–matter interactions that surpass those obtained in both top-down engineered metasurfaces and bottom-up synthesized nanocrystals. The resulting high packing density promotes strong dipole–dipole coupling, which dominates the plasmonic response of the material. Ultimately, near-field coupling elevates the size-distribution of doped nanograins to a central parameter controlling the plasmonic response, even if grain dimensions lie nearly three orders of magnitude below the excitation wavelength. Altogether, these results establish nano-polysilicon as a compelling platform for sustainable and low-cost infra-red plasmonic materials, eliminating complex nanofabrication steps while remaining fully compatible with microelectronic and photovoltaic technologies.

\section{Acknowledgments} This work was  partly supported by ANR DIAAPASON (ANR-24-CE09-4555) and by the Tremplin project PLASMONIX from CNRS-Physique.

\printbibliography

@article{faucheaux_plasmon_2014,
	title = {Plasmon {Resonances} of {Semiconductor} {Nanocrystals}: {Physical} {Principles} and {New} {Opportunities}},
	volume = {5},
	issn = {1948-7185},
	shorttitle = {Plasmon {Resonances} of {Semiconductor} {Nanocrystals}},
	url = {http://pubs.acs.org/doi/10.1021/jz500037k},
	doi = {10.1021/jz500037k},
	language = {en},
	number = {6},
	urldate = {2019-05-17},
	journal = {The Journal of Physical Chemistry Letters},
	author = {Faucheaux, Jacob A. and Stanton, Alexandria L. D. and Jain, Prashant K.},
	month = mar,
	year = {2014},
	pages = {976--985},
}

@article{luther_localized_2011,
	title = {Localized surface plasmon resonances arising from free carriers in doped quantum dots},
	volume = {10},
	issn = {1476-1122, 1476-4660},
	url = {http://www.nature.com/articles/nmat3004},
	doi = {10.1038/nmat3004},
	language = {en},
	number = {5},
	urldate = {2019-05-17},
	journal = {Nature Materials},
	author = {Luther, Joseph M. and Jain, Prashant K. and Ewers, Trevor and Alivisatos, A. Paul},
	month = may,
	year = {2011},
	pages = {361--366},
}

@article{kriegel_plasmonic_2017,
	title = {Plasmonic doped semiconductor nanocrystals: {Properties}, fabrication, applications and perspectives},
	volume = {674},
	issn = {03701573},
	shorttitle = {Plasmonic doped semiconductor nanocrystals},
	url = {https://linkinghub.elsevier.com/retrieve/pii/S0370157317300364},
	doi = {10.1016/j.physrep.2017.01.003},
	language = {en},
	urldate = {2019-05-17},
	journal = {Physics Reports},
	author = {Kriegel, Ilka and Scotognella, Francesco and Manna, Liberato},
	month = feb,
	year = {2017},
	pages = {1--52},
}

@article{taliercio_semiconductor_2019,
	title = {Semiconductor infrared plasmonics},
	volume = {8},
	issn = {2192-8614},
	url = {http://www.degruyter.com/view/j/nanoph.2019.8.issue-6/nanoph-2019-0077/nanoph-2019-0077.xml},
	doi = {10.1515/nanoph-2019-0077},
	language = {en},
	number = {6},
	urldate = {2020-03-09},
	journal = {Nanophotonics},
	author = {Taliercio, Thierry and Biagioni, Paolo},
	month = jun,
	year = {2019},
	pages = {949--990},
}

@article{pellegrini_benchmarking_2018,
	title = {Benchmarking the {Use} of {Heavily} {Doped} {Ge} for {Plasmonics} and {Sensing} in the {Mid}-{Infrared}},
	volume = {5},
	issn = {2330-4022, 2330-4022},
	url = {https://pubs.acs.org/doi/10.1021/acsphotonics.8b00438},
	doi = {10.1021/acsphotonics.8b00438},
	language = {en},
	number = {9},
	urldate = {2020-03-09},
	journal = {ACS Photonics},
	author = {Pellegrini, Giovanni and Baldassare, Leonetta and Giliberti, Valeria and Frigerio, Jacopo and Gallacher, Kevin and Paul, Douglas J. and Isella, Giovanni and Ortolani, Michele and Biagioni, Paolo},
	month = sep,
	year = {2018},
	pages = {3601--3607},
}

@article{naik_alternative_2013,
	title = {Alternative {Plasmonic} {Materials}: {Beyond} {Gold} and {Silver}},
	volume = {25},
	issn = {09359648},
	shorttitle = {Alternative {Plasmonic} {Materials}},
	url = {http://doi.wiley.com/10.1002/adma.201205076},
	doi = {10.1002/adma.201205076},
	language = {en},
	number = {24},
	urldate = {2020-03-27},
	journal = {Advanced Materials},
	author = {Naik, Gururaj V. and Shalaev, Vladimir M. and Boltasseva, Alexandra},
	month = jun,
	year = {2013},
	pages = {3264--3294},
}

@article{baldassarre_midinfrared_2015,
	title = {Midinfrared {Plasmon}-{Enhanced} {Spectroscopy} with {Germanium} {Antennas} on {Silicon} {Substrates}},
	volume = {15},
	issn = {1530-6984, 1530-6992},
	url = {https://pubs.acs.org/doi/10.1021/acs.nanolett.5b03247},
	doi = {10.1021/acs.nanolett.5b03247},
	language = {en},
	number = {11},
	urldate = {2020-09-09},
	journal = {Nano Letters},
	author = {Baldassarre, Leonetta and Sakat, Emilie and Frigerio, Jacopo and Samarelli, Antonio and Gallacher, Kevin and Calandrini, Eugenio and Isella, Giovanni and Paul, Douglas J. and Ortolani, Michele and Biagioni, Paolo},
	month = nov,
	year = {2015},
	pages = {7225--7231},
}

@book{pichler_intrinsic_2004,
	address = {Vienna},
	series = {Computational {Microelectronics}},
	title = {Intrinsic {Point} {Defects}, {Impurities}, and {Their} {Diffusion} in {Silicon}},
	isbn = {978-3-7091-7204-9 978-3-7091-0597-9},
	url = {http://link.springer.com/10.1007/978-3-7091-0597-9},
	language = {en},
	urldate = {2020-09-28},
	publisher = {Springer Vienna},
	author = {Pichler, Peter},
	editor = {Selberherr, S.},
	year = {2004},
	doi = {10.1007/978-3-7091-0597-9},
}

@article{faucheaux_plasmons_2013,
	title = {Plasmons in {Photocharged} {ZnO} {Nanocrystals} {Revealing} the {Nature} of {Charge} {Dynamics}},
	volume = {4},
	issn = {1948-7185, 1948-7185},
	url = {https://pubs.acs.org/doi/10.1021/jz401719u},
	doi = {10.1021/jz401719u},
	language = {en},
	number = {18},
	urldate = {2025-03-26},
	journal = {The Journal of Physical Chemistry Letters},
	author = {Faucheaux, Jacob A. and Jain, Prashant K.},
	month = sep,
	year = {2013},
	pages = {3024--3030},
}

@article{ding_tem_2020,
	title = {{TEM} investigation of the role of the polycrystalline-silicon film/substrate interface in high quality radio frequency silicon substrates},
	volume = {161},
	issn = {10445803},
	url = {https://linkinghub.elsevier.com/retrieve/pii/S104458031933339X},
	doi = {10.1016/j.matchar.2020.110174},
	language = {en},
	urldate = {2025-09-03},
	journal = {Materials Characterization},
	author = {Ding, Lipeng and Raskin, Jean-Pierre and Lumbeeck, Gunnar and Schryvers, Dominique and Idrissi, Hosni},
	month = mar,
	year = {2020},
	pages = {110174},
}

@book{maier_plasmonics_2007,
	address = {New York, NY},
	edition = {Repr.},
	title = {Plasmonics: fundamentals and applications},
	isbn = {978-0-387-33150-8},
	shorttitle = {Plasmonics},
	language = {en},
	publisher = {Springer},
	author = {Maier, Stefan A.},
	year = {2007},
}

@article{rauch_automated_2010,
	title = {Automated nanocrystal orientation and phase mapping in the transmission electron microscope on the basis of precession electron diffraction},
	volume = {225},
	issn = {0044-2968},
	url = {https://www.degruyterbrill.com/document/doi/10.1524/zkri.2010.1205/html},
	doi = {10.1524/zkri.2010.1205},
	language = {en},
	number = {2-3},
	urldate = {2025-11-19},
	journal = {Zeitschrift für Kristallographie},
	author = {Rauch, Edgar F. and Portillo, Joaquin and Nicolopoulos, Stavros and Bultreys, Daniel and Rouvimov, Sergei and Moeck, Peter},
	month = mar,
	year = {2010},
	pages = {103--109},
}

@article{law_towards_2013,
	title = {Towards nano-scale photonics with micro-scale photons: the opportunities and challenges of mid-infrared plasmonics},
	volume = {2},
	issn = {2192-8614, 2192-8606},
	shorttitle = {Towards nano-scale photonics with micro-scale photons},
	url = {https://www.degruyter.com/document/doi/10.1515/nanoph-2012-0027/html},
	doi = {10.1515/nanoph-2012-0027},
	language = {en},
	number = {2},
	urldate = {2025-11-19},
	journal = {Nanophotonics},
	author = {Law, Stephanie and Podolskiy, Viktor and Wasserman, Daniel},
	month = apr,
	year = {2013},
	pages = {103--130},
}

@article{huet_doping_2017,
	title = {Doping of semiconductor devices by {Laser} {Thermal} {Annealing}},
	volume = {62},
	issn = {13698001},
	url = {https://linkinghub.elsevier.com/retrieve/pii/S1369800116305017},
	doi = {10.1016/j.mssp.2016.11.008},
	language = {en},
	urldate = {2025-11-19},
	journal = {Materials Science in Semiconductor Processing},
	author = {Huet, Karim and Mazzamuto, Fulvio and Tabata, Toshiyuki and Toqué-Tresonne, Ines and Mori, Yoshihiro},
	month = may,
	year = {2017},
	pages = {92--102},
}

@article{chery_study_2022,
	title = {Study of recrystallization and activation processes in thin and highly doped silicon-on-insulator layers by nanosecond laser thermal annealing},
	volume = {131},
	issn = {0021-8979, 1089-7550},
	url = {https://pubs.aip.org/jap/article/131/6/065301/2836537/Study-of-recrystallization-and-activation},
	doi = {10.1063/5.0073827},
	language = {en},
	number = {6},
	urldate = {2025-09-18},
	journal = {Journal of Applied Physics},
	author = {Chery, N. and Zhang, M. and Monflier, R. and Mallet, N. and Seine, G. and Paillard, V. and Poumirol, J. M. and Larrieu, G. and Royet, A. S. and Kerdilès, S. and Acosta-Alba, P. and Perego, M. and Bonafos, C. and Cristiano, F.},
	month = feb,
	year = {2022},
	pages = {065301},
}

@article{kramer_plasmonic_2015,
	title = {Plasmonic {Properties} of {Silicon} {Nanocrystals} {Doped} with {Boron} and {Phosphorus}},
	volume = {15},
	issn = {1530-6984, 1530-6992},
	url = {http://pubs.acs.org/doi/10.1021/acs.nanolett.5b02287},
	doi = {10.1021/acs.nanolett.5b02287},
	language = {en},
	number = {8},
	urldate = {2019-05-17},
	journal = {Nano Letters},
	author = {Kramer, Nicolaas J. and Schramke, Katelyn S. and Kortshagen, Uwe R.},
	month = aug,
	year = {2015},
	pages = {5597--5603},
}

@article{poumirol_hyper-doped_2021,
	title = {Hyper-{Doped} {Silicon} {Nanoantennas} and {Metasurfaces} for {Tunable} {Infrared} {Plasmonics}},
	volume = {8},
	copyright = {https://doi.org/10.15223/policy-029},
	issn = {2330-4022, 2330-4022},
	url = {https://pubs.acs.org/doi/10.1021/acsphotonics.1c00019},
	doi = {10.1021/acsphotonics.1c00019},
	language = {en},
	number = {5},
	urldate = {2025-09-18},
	journal = {ACS Photonics},
	author = {Poumirol, Jean-Marie and Majorel, Clément and Chery, Nicolas and Girard, Christian and Wiecha, Peter R. and Mallet, Nicolas and Monflier, Richard and Larrieu, Guilhem and Cristiano, Filadelfo and Royet, Anne-Sophie and Alba, Pablo Acosta and Kerdiles, Sébastien and Paillard, Vincent and Bonafos, Caroline},
	month = may,
	year = {2021},
	pages = {1393--1399},
}

@article{valdenaire_heavily_2023,
	title = {Heavily {Doped} {Si} {Nanocrystals} {Formed} in {P}-({SiO}/{SiO}$_{\textrm{2}}$ ) {Multilayers}: {A} {Promising} {Route} for {Si}-{Based} {Infrared} {Plasmonics}},
	volume = {6},
	copyright = {https://doi.org/10.15223/policy-029},
	issn = {2574-0970, 2574-0970},
	shorttitle = {Heavily {Doped} {Si} {Nanocrystals} {Formed} in {P}-({SiO}/{SiO}$_{\textrm{2}}$ ) {Multilayers}},
	url = {https://pubs.acs.org/doi/10.1021/acsanm.2c05088},
	doi = {10.1021/acsanm.2c05088},
	language = {en},
	number = {5},
	urldate = {2025-09-18},
	journal = {ACS Applied Nano Materials},
	author = {Valdenaire, Alix and Giba, Alaa Eldin and Stoffel, Mathieu and Devaux, Xavier and Foussat, Loïc and Poumirol, Jean-Marie and Bonafos, Caroline and Guehairia, Sonia and Demoulin, Rémi and Talbot, Etienne and Vergnat, Michel and Rinnert, Hervé},
	month = mar,
	year = {2023},
	pages = {3312--3320},
}

@article{zhang_infrared_2022,
	title = {Infrared nanoplasmonic properties of hyperdoped embedded {Si} nanocrystals in the few electrons regime},
	volume = {11},
	copyright = {http://creativecommons.org/licenses/by/4.0},
	issn = {2192-8614},
	url = {https://www.degruyter.com/document/doi/10.1515/nanoph-2022-0283/html},
	doi = {10.1515/nanoph-2022-0283},
	language = {en},
	number = {15},
	urldate = {2025-09-18},
	journal = {Nanophotonics},
	author = {Zhang, Meiling and Poumirol, Jean-Marie and Chery, Nicolas and Majorel, Clément and Demoulin, Rémi and Talbot, Etienne and Rinnert, Hervé and Girard, Christian and Cristiano, Fuccio and Wiecha, Peter R. and Hungria, Teresa and Paillard, Vincent and Arbouet, Arnaud and Pécassou, Béatrice and Gourbilleau, Fabrice and Bonafos, Caroline},
	month = aug,
	year = {2022},
	pages = {3485--3493},
}

@article{zhou_comparative_2015,
	title = {Comparative {Study} on the {Localized} {Surface} {Plasmon} {Resonance} of {Boron}- and {Phosphorus}-{Doped} {Silicon} {Nanocrystals}},
	volume = {9},
	issn = {1936-0851},
	url = {https://doi.org/10.1021/nn505416r},
	doi = {10.1021/nn505416r},
	number = {1},
	urldate = {2020-03-09},
	journal = {ACS Nano},
	author = {Zhou, Shu and Pi, Xiaodong and Ni, Zhenyi and Ding, Yi and Jiang, Yingying and Jin, Chuanhong and Delerue, Christophe and Yang, Deren and Nozaki, Tomohiro},
	month = jan,
	year = {2015},
	note = {Publisher: American Chemical Society},
	pages = {378--386},
}

@article{rechberger_optical_2003,
	title = {Optical properties of two interacting gold nanoparticles},
	volume = {220},
	copyright = {https://www.elsevier.com/tdm/userlicense/1.0/},
	issn = {00304018},
	url = {https://linkinghub.elsevier.com/retrieve/pii/S0030401803013579},
	doi = {10.1016/S0030-4018(03)01357-9},
	language = {en},
	number = {1-3},
	urldate = {2025-12-09},
	journal = {Optics Communications},
	author = {Rechberger, W. and Hohenau, A. and Leitner, A. and Krenn, J.R. and Lamprecht, B. and Aussenegg, F.R.},
	month = may,
	year = {2003},
	pages = {137--141},
}

@article{su_interparticle_2003,
	title = {Interparticle {Coupling} {Effects} on {Plasmon} {Resonances} of {Nanogold} {Particles}},
	volume = {3},
	issn = {1530-6984, 1530-6992},
	url = {https://pubs.acs.org/doi/10.1021/nl034197f},
	doi = {10.1021/nl034197f},
	language = {en},
	number = {8},
	urldate = {2025-12-09},
	journal = {Nano Letters},
	author = {Su, K.-H. and Wei, Q.-H. and Zhang, X. and Mock, J. J. and Smith, D. R. and Schultz, S.},
	month = aug,
	year = {2003},
	pages = {1087--1090},
}

@article{maier_observation_2002,
	title = {Observation of near-field coupling in metal nanoparticle chains using far-field polarization spectroscopy},
	volume = {65},
	copyright = {http://link.aps.org/licenses/aps-default-license},
	issn = {0163-1829, 1095-3795},
	url = {https://link.aps.org/doi/10.1103/PhysRevB.65.193408},
	doi = {10.1103/PhysRevB.65.193408},
	language = {en},
	number = {19},
	urldate = {2025-12-10},
	journal = {Physical Review B},
	author = {Maier, Stefan A. and Brongersma, Mark L. and Kik, Pieter G. and Atwater, Harry A.},
	month = may,
	year = {2002},
	pages = {193408},
}

@article{gunnarsson_confined_2005,
	title = {Confined {Plasmons} in {Nanofabricated} {Single} {Silver} {Particle} {Pairs}: {Experimental} {Observations} of {Strong} {Interparticle} {Interactions}},
	volume = {109},
	issn = {1520-6106, 1520-5207},
	shorttitle = {Confined {Plasmons} in {Nanofabricated} {Single} {Silver} {Particle} {Pairs}},
	url = {https://pubs.acs.org/doi/10.1021/jp049084e},
	doi = {10.1021/jp049084e},
	language = {en},
	number = {3},
	urldate = {2025-12-16},
	journal = {The Journal of Physical Chemistry B},
	author = {Gunnarsson, Linda and Rindzevicius, Tomas and Prikulis, Juris and Kasemo, Bengt and Käll, Mikael and Zou, Shengli and Schatz, George C.},
	month = jan,
	year = {2005},
	pages = {1079--1087},
}

@article{sakaguchi_atomic_2007,
	title = {Atomic {Structure} of {Faceted} \&{Sigma};3 {CSL} {Grain} {Boundary} in {Silicon}: {HRTEM} and {\textless}{I}{\textgreater}{Ab}-initio{\textless}/{I}{\textgreater} {Calculation}},
	volume = {48},
	issn = {1345-9678, 1347-5320},
	shorttitle = {Atomic {Structure} of {Faceted} \&{Sigma};3 {CSL} {Grain} {Boundary} in {Silicon}},
	url = {https://www.jstage.jst.go.jp/article/matertrans/48/10/48_MD200706/_article},
	doi = {10.2320/matertrans.MD200706},
	language = {en},
	number = {10},
	urldate = {2025-12-18},
	journal = {MATERIALS TRANSACTIONS},
	author = {Sakaguchi, Norihito and Ichinose, Hideki and Watanabe, Seiichi},
	year = {2007},
	pages = {2585--2589},
}

@article{kim_theory_1984,
	title = {Theory of conduction in polysilicon: {Drift}-diffusion approach in crystalline-amorphous-crystalline semiconductor system\&\#8212;{Part} {I}: {Small} signal theory},
    year={1984}, 
	volume = {31},
	copyright = {https://ieeexplore.ieee.org/Xplorehelp/downloads/license-information/IEEE.html},
	issn = {0018-9383, 1557-9646},
	shorttitle = {Theory of conduction in polysilicon},
	url = {https://ieeexplore.ieee.org/document/1483839/},
	doi = {10.1109/T-ED.1984.21554},
	language = {en},
	number = {4},
	urldate = {2025-12-18},
	journal = {IEEE Transactions on Electron Devices},
	author = {Kim, D.M. and Khondker, A.N. and Ahmed, S.S. and Shah, R.R.},
	month = apr,
	year = {1984},
	pages = {480--493},
}

@Inbook{Kamins1998,
author="Kamins, Ted",
title="Electrical Properties",
bookTitle="Polycrystalline Silicon for Integrated Circuits and Displays",
year="1998",
publisher="Springer US",
address="Boston, MA",
pages="195--243",
isbn="978-1-4615-5577-3",
doi="10.1007/978-1-4615-5577-3_5",
url="https://doi.org/10.1007/978-1-4615-5577-3_5"
}

@article{yang_plasmonic_2025,
	title = {Plasmonic metasurfaces: {Light}-matter interactions, fabrication, applications and future outlooks},
	volume = {154},
	issn = {00796425},
	shorttitle = {Plasmonic metasurfaces},
	url = {https://linkinghub.elsevier.com/retrieve/pii/S0079642525000866},
	doi = {10.1016/j.pmatsci.2025.101508},
	language = {en},
	urldate = {2026-01-09},
	journal = {Progress in Materials Science},
	author = {Yang, Fan and Cao, Wei and Zheng, Guangchao and Qiu, Li and Nie, Zhihong and Li, Yue},
	month = nov,
	year = {2025},
	pages = {101508},
}

@article{seong_cost-effective_2024,
	title = {Cost-{Effective} and {Environmentally} {Friendly} {Mass} {Manufacturing} of {Optical} {Metasurfaces} {Towards} {Practical} {Applications} and {Commercialization}},
	volume = {11},
	issn = {2288-6206, 2198-0810},
	url = {https://link.springer.com/10.1007/s40684-023-00580-x},
	doi = {10.1007/s40684-023-00580-x},
	language = {en},
	number = {2},
	urldate = {2026-01-09},
	journal = {International Journal of Precision Engineering and Manufacturing-Green Technology},
	author = {Seong, Junhwa and Jeon, Youngsun and Yang, Younghwan and Badloe, Trevon and Rho, Junsuk},
	month = mar,
	year = {2024},
	pages = {685--706},
}

@article{graedel_criticality_2015,
	title = {Criticality of metals and metalloids},
	volume = {112},
	issn = {0027-8424, 1091-6490},
	url = {https://pnas.org/doi/full/10.1073/pnas.1500415112},
	doi = {10.1073/pnas.1500415112},
	language = {en},
	number = {14},
	urldate = {2026-01-09},
	journal = {Proceedings of the National Academy of Sciences},
	author = {Graedel, T. E. and Harper, E. M. and Nassar, N. T. and Nuss, Philip and Reck, Barbara K.},
	month = apr,
	year = {2015},
	pages = {4257--4262},
}

@article{wang_plasmonic_2016,
	title = {Plasmonic and {Dielectric} {Metasurfaces}: {Design}, {Fabrication} and {Applications}},
	volume = {6},
	issn = {2076-3417},
	shorttitle = {Plasmonic and {Dielectric} {Metasurfaces}},
	url = {https://www.mdpi.com/2076-3417/6/9/239},
	doi = {10.3390/app6090239},
	language = {en},
	number = {9},
	urldate = {2026-01-09},
	journal = {Applied Sciences},
	author = {Wang, Jian and Du, Jing},
	month = sep,
	year = {2016},
	pages = {239},

}

@article{shahzad_infrared_2011,
	title = {Infrared surface plasmons on heavily doped silicon},
	volume = {110},
	issn = {0021-8979, 1089-7550},
	url = {https://pubs.aip.org/jap/article/110/12/123105/698970/Infrared-surface-plasmons-on-heavily-doped-silicon},
	doi = {10.1063/1.3672738},
	language = {en},
	number = {12},
	urldate = {2026-01-12},
	journal = {Journal of Applied Physics},
	author = {Shahzad, Monas and Medhi, Gautam and Peale, Robert E. and Buchwald, Walter R. and Cleary, Justin W. and Soref, Richard and Boreman, Glenn D. and Edwards, Oliver},
	month = dec,
	year = {2011},
	pages = {123105},
}

@article{faruque_heavily_2021,
	title = {Heavily doped silicon: {A} potential replacement of conventional plasmonic metals},
	volume = {42},
	issn = {1674-4926, 2058-6140},
	shorttitle = {Heavily doped silicon},
	url = {https://iopscience.iop.org/article/10.1088/1674-4926/42/6/062302},
	doi = {10.1088/1674-4926/42/6/062302},
	language = {en},
	number = {6},
	urldate = {2025-09-03},
	journal = {Journal of Semiconductors},
	author = {Faruque, Md. Omar and Al Mahmud, Rabiul and Sagor, Rakibul Hasan},
	month = jun,
	year = {2021},
	pages = {062302},
}

@article{tsurekawa_interfacial_2007,
	title = {Interfacial state and potential barrier height associated with grain boundaries in polycrystalline silicon},
	volume = {462},
	copyright = {https://www.elsevier.com/tdm/userlicense/1.0/},
	issn = {09215093},
	url = {https://linkinghub.elsevier.com/retrieve/pii/S092150930601851X},
	doi = {10.1016/j.msea.2006.02.471},
	language = {en},
	number = {1-2},
	urldate = {2026-01-14},
	journal = {Materials Science and Engineering: A},
	author = {Tsurekawa, Sadahiro and Kido, Kota and Watanabe, Tadao},
	month = jul,
	year = {2007},
	pages = {61--67},
}

@article{mandurah_model_1981,
	title = {A model for conduction in polycrystalline silicon\&\#8212;{Part} {I}: {Theory}},
	volume = {28},
	copyright = {https://ieeexplore.ieee.org/Xplorehelp/downloads/license-information/IEEE.html},
	issn = {0018-9383, 1557-9646},
	shorttitle = {A model for conduction in polycrystalline silicon\&\#8212;{Part} {I}},
	url = {https://ieeexplore.ieee.org/document/1481656/},
	doi = {10.1109/T-ED.1981.20504},
	language = {en},
	number = {10},
	urldate = {2026-01-14},
	journal = {IEEE Transactions on Electron Devices},
	author = {Mandurah, M.M. and Saraswat, K.C. and Kamins, T.I.},
	month = oct,
	year = {1981},
	pages = {1163--1171},
}

@article{hu_review_2021,
	title = {A {Review} on {Metasurface}: {From} {Principle} to {Smart} {Metadevices}},
	volume = {8},
	issn = {2296-424X},
	shorttitle = {A {Review} on {Metasurface}},
	url = {https://www.frontiersin.org/articles/10.3389/fphy.2020.586087/full},
	doi = {10.3389/fphy.2020.586087},
	language = {en},
	urldate = {2026-01-14},
	journal = {Frontiers in Physics},
	author = {Hu, Jie and Bandyopadhyay, Sankhyabrata and Liu, Yu-hui and Shao, Li-yang},
	month = jan,
	year = {2021},
	pages = {586087},
}

@article{cortes_optical_2022,
	title = {Optical {Metasurfaces} for {Energy} {Conversion}},
	volume = {122},
	copyright = {https://creativecommons.org/licenses/by-nc-nd/4.0/},
	issn = {0009-2665, 1520-6890},
	url = {https://pubs.acs.org/doi/10.1021/acs.chemrev.2c00078},
	doi = {10.1021/acs.chemrev.2c00078},
	language = {en},
	number = {19},
	urldate = {2026-01-14},
	journal = {Chemical Reviews},
	author = {Cortés, Emiliano and Wendisch, Fedja J. and Sortino, Luca and Mancini, Andrea and Ezendam, Simone and Saris, Seryio and De S. Menezes, Leonardo and Tittl, Andreas and Ren, Haoran and Maier, Stefan A.},
	month = oct,
	year = {2022},
	pages = {15082--15176},
}

@article{meinzer_plasmonic_2014,
	title = {Plasmonic meta-atoms and metasurfaces},
	volume = {8},
	issn = {1749-4885, 1749-4893},
	url = {https://www.nature.com/articles/nphoton.2014.247},
	doi = {10.1038/nphoton.2014.247},
	language = {en},
	number = {12},
	urldate = {2026-01-15},
	journal = {Nature Photonics},
	author = {Meinzer, Nina and Barnes, William L. and Hooper, Ian R.},
	month = dec,
	year = {2014},
	pages = {889--898},
}

@article{oskooi2010meep,
  title={MEEP: A flexible free-software package for electromagnetic simulations by the FDTD method},
  author={Oskooi, Ardavan F and Roundy, David and Ibanescu, Mihai and Bermel, Peter and Joannopoulos, John D and Johnson, Steven G},
  journal={Computer Physics Communications},
  volume={181},
  number={3},
  pages={687--702},
  year={2010},
  publisher={Elsevier}
}

@article{bergmann_formation_1998,
	title = {Formation of polycrystalline silicon with log-normal grain size distribution},
	volume = {123-124},
	copyright = {https://www.elsevier.com/tdm/userlicense/1.0/},
	issn = {01694332},
	url = {https://linkinghub.elsevier.com/retrieve/pii/S0169433297004947},
	doi = {10.1016/S0169-4332(97)00494-7},
	language = {en},
	urldate = {2026-02-24},
	journal = {Applied Surface Science},
	author = {Bergmann, Ralf B. and Shi, Frank G. and Queisser, Hans J. and Krinke, Jörg},
	month = jan,
	year = {1998},
	pages = {376--380}}

@article{patoux_challenges_2021,
	title = {Challenges in nanofabrication for efficient optical metasurfaces},
	volume = {11},
	issn = {2045-2322},
	url = {https://www.nature.com/articles/s41598-021-84666-z},
	doi = {10.1038/s41598-021-84666-z},
	language = {en},
	number = {1},
	urldate = {2026-02-24},
	journal = {Scientific Reports},
	author = {Patoux, Adelin and Agez, Gonzague and Girard, Christian and Paillard, Vincent and Wiecha, Peter R. and Lecestre, Aurélie and Carcenac, Franck and Larrieu, Guilhem and Arbouet, Arnaud},
	month = mar,
	year = {2021},
	pages = {5620},
}

@article{zhang_hyperdoped_2023,
	title = {Hyperdoped {Si} nanocrystals embedded in silica for infrared plasmonics},
	volume = {15},
	issn = {2040-3364, 2040-3372},
	url = {https://xlink.rsc.org/?DOI=D3NR00035D},
	doi = {10.1039/D3NR00035D},
	language = {en},
	number = {16},
	urldate = {2025-09-18},
	journal = {Nanoscale},
	author = {Zhang, Meiling and Poumirol, Jean-Marie and Chery, Nicolas and Rinnert, Hervé and Giba, Alaa E. and Demoulin, Rémi and Talbot, Etienne and Cristiano, Fuccio and Hungria, Teresa and Paillard, Vincent and Gourbilleau, Fabrice and Bonafos, Caroline},
	year = {2023},
	pages = {7438--7449},
}

\end{document}